\documentclass[a4paper, reprint, groupedaddress, floatfix, superscriptaddress, showpacs, prl, cha]{revtex4-1}

\usepackage{amssymb}
\usepackage{amsmath}

\usepackage{dcolumn}
\usepackage{subfigure}
\usepackage[dvips]{graphicx}
\usepackage[english]{babel}
\usepackage[usenames]{color}
\usepackage[]{varioref}
\usepackage[latin1]{inputenc}
\usepackage{docs}
\usepackage{bm}
\usepackage{booktabs}

\begin{document}

\title{Tunable spin pumping in exchange coupled magnetic trilayers}

\author{M. Fazlali}
\affiliation{Department of Physics, University of Gothenburg, 412 96, Gothenburg, Sweden}

\author{M. Ahlberg}
\affiliation{Department of Physics, University of Gothenburg, 412 96, Gothenburg, Sweden}

\author{M. Dvornik}
\affiliation{Department of Physics, University of Gothenburg, 412 96, Gothenburg, Sweden}

\author{J. \AA{}kerman}
\affiliation{Department of Physics, University of Gothenburg, 412 96, Gothenburg, Sweden}
\affiliation{Department of Materials and Nano Physics, School of Information and Communication Technology, KTH Royal Institute of Technology, Electrum 229, 164 40 Kista, Sweden}

\date{\today}

\begin{abstract}
Magnetic thin films at ferromagnetic resonance (FMR) leak angular momentum, which may be absorbed by adjacent layers. This phenomenon, known as spin pumping, is manifested by an increase in the resonance linewidth ($\Delta H$), and the closely related Gilbert damping. Another effect of this transfer of spin currents is a dynamical and long-range coupling that can drive two magnetic layers into a collective precession when their FMR frequencies coincide. A collective behavior is also found in magnetic trilayers with interlayer exchange coupling (IEC). In this study we investigate the interplay between IEC and spin pumping, using Co/Cu/Py pseudo-spin values. We employ broadband FMR spectroscopy to explore both the frequency and coupling-strength dependence of $\Delta H$. Our observations show that there exists a cut-off frequency, set by the IEC strength, below which the precession is truly collective and the spin pumping is suppressed.  These results demonstrate that it is possible to control the spin pumping efficiency by varying the frequency or the interlayer exchange coupling.
\end{abstract}

\maketitle 

Pseudo-spin valves are the building blocks of many spintronic devices, such as nanocontact spin-torque nano-oscillators~\cite{Tsoi1998prl,Slonczewski1999jmmm,Silva2008jmmm,Slavin2009ieeetmag,Dumas2014ieeetmag,Chen2016}. These devices not only show great promise for microwave~\cite{Bonetti2009apl,Banuazizi2017} and magnonic~\cite{Bonetti2013tap} applications, but have also allowed for the exploration of spin transfer torque driven propagating spin waves~\cite{Madami2011nn,Bonetti2010prl} and magnetodynamical solitons~\cite{Slavin2005prl,Bonetti2010prl,Demidov2010nm,Dumas2013, Mohseni2013}. Two cornerstones for further development towards applications, and to open new routes to answer fundamental questions are to tailor the magnetic damping and to control the flow of spin currents. The concept of spin pumping describes how the leakage of angular momentum (spin current) from a precessing magnetic film may be absorbed at the interface to another magnetic/non-magnetic layer, which provides an additional damping term~\cite{Urban2001, Tserkovnyak2002, Heinrich2003a}. The dimensionless damping coefficient is then given by $\alpha = \alpha_{(0)} + \alpha_{\mathrm{sp}}$, where $\alpha_{(0)}$ is the intrinsic damping of the precessing layer and $\alpha_{\mathrm{sp}}$ is the spin-pumping-induced term. While this effect has been studied extensively~\cite{Tserkovnyak2005}, the majority of those investigation has focused on the regime where the static coupling between the layers is very weak.  

The static interlayer exchange coupling (IEC, $J'$), is an oscillatory coupling present when two ferromagnetic films (FMs) are separated by a sufficiently thin nonmagnetic layer (NM)~\cite{Parkin1990, Heinrich2008}. The coupling will either promote a parallel or antiparallel configuration of the magnetization in the films, depending on the spacer layer thickness. It can be described within the Ruderman-Kittel-Kasuya-Yosida framework~\cite{Bruno1991} and is therefore often referred to as RKKY-interaction. For sufficiently strong $J'$, the IEC can separate the ferromagnetic resonance (FMR) of a FM/NM/FM trilayer into two modes, called the acoustic (in-phase) and optical (out-of-phase) mode~\cite{Heinrich1994}. While the static IEC is oscillating and short-ranged in nature, there also exists a dynamic and long-ranged coupling between magnetic layers. Two magnetic films that precess at the same frequency can synchronize and display a collective behavior in the presence of spin pumping~\cite{Heinrich2003a, Heinrich2008}. The exploration of the interplay between the static and dynamic interlayer exchange coupling hence has great prospects of finding new exciting physics. 

While the literature includes studies on spin pumping in coupled layers~\cite{Lenz2004, Lenz2004b, Timopheev2014, Baker2016a, Baker2016}, most focus on one or a few frequencies. Here we investigate spin pumping in a wide frequency range of 3-37~GHz using a broadband FMR setup and examine four regimes: strong, intermediate, weak, and zero IEC. The samples are based on Co/Cu/Py trilayers, where the thickness of the Cu spacer sets the strength of the interlayer interaction. 

We show that the mode hybridization between the layers, leading to acoustical and optical modes, is not only dependent on the IEC but also on the field and frequency. The collective nature of the precession is clear at low fields, as reflected by the relative amplitude of the signals and the field dependence of the resonance frequency  ($f_{\mathrm{r}}$). At higher applied fields the layers instead behave as single films subjected to an effective field, which scales with the interlayer coupling. This transition, from collective to single layer precession, is accompanied by changes in the slope of $\Delta H$ vs.~$f_{\mathrm{r}}$, i.e.~the damping, and we attribute those changes to the spin pumping between the layers. The results demonstrate that it is possible to engineer a \emph{cut-off frequency}, below which the spin pumping is effectively turned off. At higher frequencies the spin pumping and the concomitant damping gradually increases until it reach a constant value. This effect can be used to tailor the behavior of spintronic devices by the strength of the IEC.

The samples were prepared on oxidized Si-substrates using magnetron sputtering and have the following structure: substrate/seed/Co(80~\AA)/Cu($d_{\mathrm{Cu}}$)/Py(45~\AA)/cap, where $d_{\mathrm{Cu}}=0-40$~\AA. Single Py (Ni$_{80}$Fe$_{20}$) and Co films were also prepared, using the same seed and cap layers. The seed layer consisted of Pd(80~\AA)/Cu(150~\AA) and the cap layer was Cu(30~\AA)/Pd(30~\AA). These layers were included in the sample structure for three reasons: \emph{i}) to facilitate the growth of the Co film, \emph{ii}) to avoid oxidation, and \emph{iii}) to stay close to a material stack commonly used in the fabrication of spin-torque nano-oscillators. 

Using Py and Co with well separated resonance frequencies also allows us to observe both the acoustic and optical modes in the coupled regimes -- identical layers only display one resonance~\cite{Heinrich1994}. Moreover, by drawing the analogy with classical harmonic oscillators, we expect the collective modes to gradually split from the free running resonances of the individual layers (see, e.g., Section 2.2 in~\cite{Joe2006}). Thus, for weak coupling, the acoustical and optical modes will have their intensities mostly concentrated in the Py and Co layer, respectively. As the coupling increases, the collective modes span more over both layers, which leads to an enhancement of the acoustical mode net response and a suppression of the optical mode intensity.
The measurements were done using a NanOsc Instruments PhaseFMR-40 broadband FMR spectrometer. The external in-plane field was swept at fixed frequencies, varied step-wise from 3 to 37~GHz, and the acquired instrument signal represents the derivative of the absorption peaks. The derivative of an asymmetric Lorentzian~\cite{ Woltersdorf2001, Yin2015} was fit to the signal, providing the resonance field ($\mu_{0} H_{\mathrm{r}}$), the full width half maximum ($\Delta H$) and the amplitude ($A$) of the absorption peak. The Kittel equation~\cite{Kittel1948} was subsequently fit to the resonance frequencies of the single layer samples:
\begin{equation}
f_{\mathrm{r}}=\frac{\gamma \mu_{0}}{2 \pi} \sqrt{(H_{\mathrm{r}}+H_{\mathrm{add}})(H_{\mathrm{r}}+H_{\mathrm{add}}+M_{\mathrm{eff}})}
\label{eq:Kittel}
\end{equation}
\noindent where $\mu_{0}$ is the permabillity of free space, $\gamma$ is the gyromagnetic ratio, $M_{\mathrm{eff}}$ is the effective magnetization, $H_{\mathrm{r}}$ is the (applied) resonance field, and $H_{\mathrm{add}}$ is the sum of additional in-plane fields, mainly represented by the anisotropy. 

The results of the fits gave the following magnetic properties of Py: $\gamma/2\pi$=29.0~GHz/T and $\mu_{0} M_{\mathrm{eff}}$=0.89~T, and of Co: $\gamma/2\pi$=30.8~GHz/T and $\mu_{0} M_{\mathrm{eff}}$=1.56~T. The additional fields are in the order of 2~mT for both samples. There are examples in the literature where the Kittel equation has also been used to fit the FMR of exchange coupled layers, and it has been claimed that the magnitude of the IEC can be extracted from the fitted value of $H_{\mathrm{add}}$~\cite{Baker2016a, Stenning2015}. However, we observed that this method not only gave quite poor fits, but the $J'$ values determined independently from $H^{\mathrm{Py}}_{\mathrm{add}}$ and $H^{\mathrm{Co}}_{\mathrm{add}}$ also differed significantly. 

Instead, we used an approach where the relation between $f_{\mathrm{r}}$ and $H_{\mathrm{r}}$ is derived from the free energy of the system, giving the following expression~\cite{Lindner2003, Layadi2002, Wei2014}:
\begin{equation}
a \omega^{4} +  c \omega^{2}  + e \omega = 0
\label{eq:J}
\end{equation}
\noindent where $\omega=2 \pi f_{\mathrm{r}}$, and the coefficients $a$, $c$, and $e$ contain the interlayer coupling, the magnetic properties, as well as the thickness of the magnetic layers. We have followed the equations presented in Ref.~\cite{Layadi2002} and~\cite{Wei2014}, and the results show a clear correspondence with the data. 

\begin{figure}[t]
\centering
\includegraphics[width=8cm]{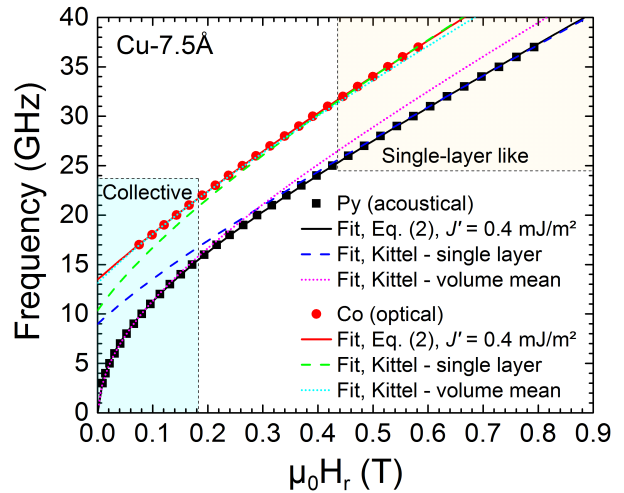}
\caption{FMR frequency vs. field for a trilayer with a 7.5~\AA~Cu spacer. The data (filled symbols) is compared to three different models. The behavior at all fields is well described by a fit of Eq.~(\ref{eq:J}) to the data, using $J'=0.4$~mJ/m$^{2}$ (solid lines). The transition from collective to single-layer-like precession is illustrated by two fits using the Kittel equation, Eq.~(\ref{eq:Kittel}), where $M_{\mathrm{S}}$ and  $\gamma$ are fixed either to the values of single Py and Co (dashed lines), or to the volume mean of these layers (dotted lines). The additional field ($H_{\mathrm{add}}$) was used as a free parameter in those fits.}
\label{fig:FreqVsField}
\end{figure}

Nevertheless, the Kittel equation still sheds some light on the nature of the oscillations. In Fig.~\ref{fig:FreqVsField} the fits of Eq.~(\ref{eq:J}) are therefore supplemented with calculations using the Kittel equation. At high fields, $f_{\mathrm{r}}$ follows the predictions of Eq.~(\ref{eq:Kittel}) with magnetizations and gyromagnetic ratios equal to those of the single films; at low fields, $f_{\mathrm{r}}$ instead matches the behavior of an effective medium with an $M_{\mathrm{eff}}$ and $\gamma$ given by the volume average of the two materials. These limiting cases hence imply a transition from a high $f_{\mathrm{r}}$ region, where the inherent properties of each layer dominates, to a low $f_{\mathrm{r}}$ region governed by collective motion.

\begin{figure}[t]
\centering
\includegraphics[width=8cm]{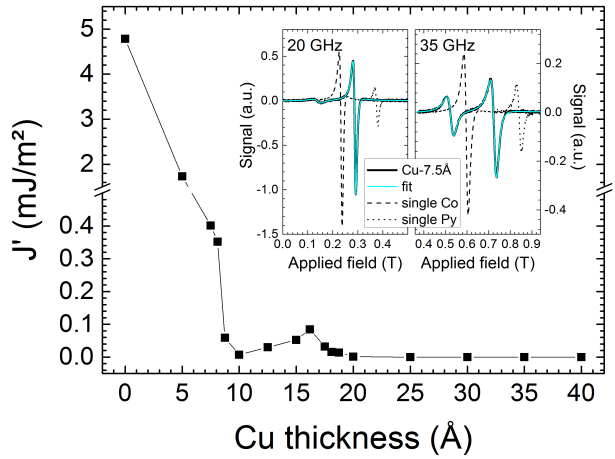}
\caption{IEC strength ($J'$) vs.~spacer layer thickness, as determined from fits of the field dependence of $f_{\mathrm{r}}$ to Eq.~(\ref{eq:J}). Inset) FMR spectra of the Cu-7.5\AA~sample in Fig.1 (thick black lines) together with fits of two asymmetric Lorentzians (cyan lines), and the spectra of the single layer Co (dashed lines) and Py (dotted lines). The left and right plots show the absorption at 20 and 35~GHz, respectively.}
\label{fig:IEC}
\end{figure}

The extracted values of the IEC are presented in Fig.~\ref{fig:IEC}. The layer thicknesses and magnetizations were fixed during the fits to Eq.~(\ref{eq:J}), while $J'$, $\gamma_{\mathrm{Py}}$, and  $\gamma_{\mathrm{Co}}$ were allowed to vary. The resulting values of the IEC were essentially unchanged if the gyromagnetic ratios were treated as constants, but the goodness-of-fit was worse. The coupling is ferromagnetic for all $t_{\mathrm{Cu}}$ in contrast to the familiar behavior where $J'$ is expected to oscillate between positive and negative values.  We can therefore conclude that the interactions between the layers are not only given by RKKY contributions, but also includes e.g., N\'{e}el (orange peel) coupling~\cite{Heinrich2008}. This interpretation is strengthened by the fact that we observe a minimum at the thickness (10~\AA) where a negative maximum should occur~\cite{Bloemen1994, Lucinski1997}.

\begin{figure}[t]
\centering
\includegraphics[width=8cm]{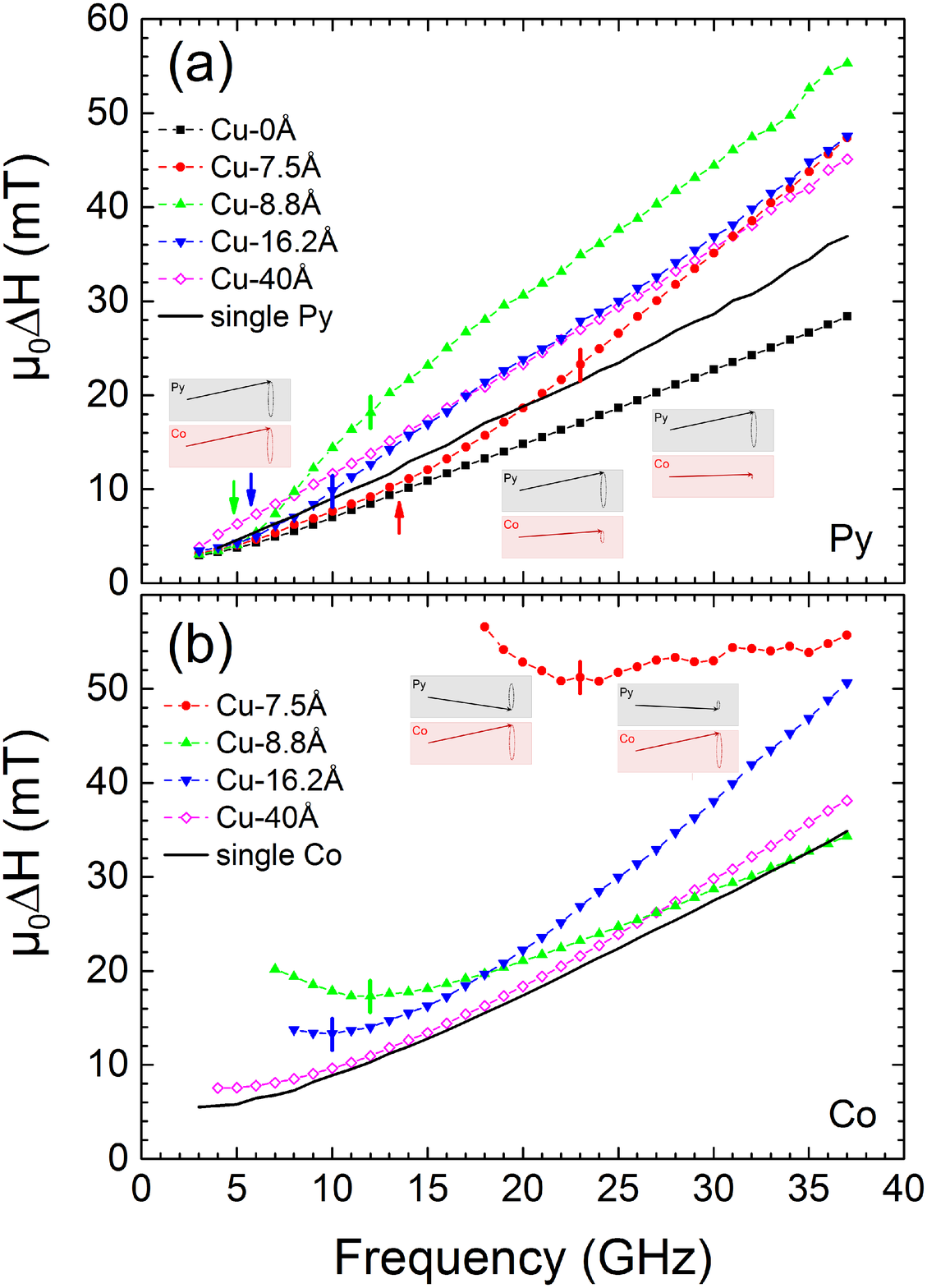}
\caption{$\Delta H$ vs.~$f_{\mathrm{r}}$ for the (a) Py and b) Co resonances. The vertical lines (in both (a) and (b)) mark the inflection point of the Co linewidth, which is marks the transition from collective to single-layer-like behavior. The arrows show the Co frequency at zero applied field. Below this frequency there is only one resonance in the system. The insets are schematic illustrations of the magnetodynamics in the Py (gray) and Co (red) layers at different frequencies, for intermediate IEC (Cu-7.5\AA).}
\label{fig:damping}
\end{figure}

The linewidth of the Py and Co resonances for representative samples are shown in Fig.~\ref{fig:damping}(a) and (b), respectively. The Cu-0\AA~and the Cu-0.5\AA~samples have strong IEC and display only the acoustic mode, while the net response of their optical mode is suppressed. The frequency dependence of the linewidth is linear at all frequencies. For samples in the intermediate and weak coupling regimes (Cu-7.5--8.8\AA; Cu-12.5--19\AA) two resonances are observed and $\Delta H$ vs.~$f_{\mathrm{r}}$ have a rather concave shape, which is more pronounced for Co. The linewidth of the single Co layer is also non-Gilbert-like, as it saturates at a constant value at low $f_{\mathrm{r}}$. This behavior is inherited in the trilayer samples. However, $\Delta H_{\mathrm{Co}}$ of the coupled layers not only flattens out, but also increases at low frequencies. This implies that the mode becomes more optical-like, since the optical mode is expected to have a much larger linewidth compared to a single layer due to mutual, out-of-phase, spin pumping~\cite{Lenz2004b,Takahashi2014, Heinrich2003b}. There is also some additional variation of $\alpha_{\mathrm{Co}}$ vs.~Cu thickness, probably due to strain induced effects beyond our control. We have consequently chosen to not dwell on the linewidth and damping associated with Co, but do note that a large $\Delta H$ is consistent with an optical character of the mode.

\begin{figure}[hbt]
\centering
\includegraphics[width=8cm]{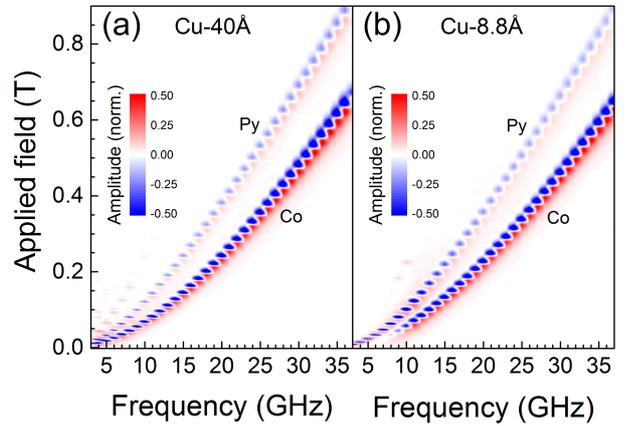}
\caption{Color map of the normalized FMR response for (a) a sample with zero, and (b) weak, interlayer coupling. The data broadening along the x-axis is an artifact arising from the conversion of line scans to a field/frequency/amplitude matrix. (a) For decoupled layers, the Py amplitude is lower than that of Co at all frequencies. (b) With IEC the relative intensities change significantly below 15~GHz. The modification of the intensities mirrors the transition from single-layer like behavior at high frequencies/fields, to a region where the precession is truly collective with acoustic (high amplitude) and optical (low amplitude) modes.}
\label{fig:contour}
\end{figure}

If we follow $\Delta H_{\mathrm{Py}}$ of the samples with intermediate/weak IEC from high to low frequencies, we see that the initial constant slope is reduced at a certain frequency ($f_{\mathrm{infl}}$) marked by a vertical line in Fig.~\ref{fig:damping}(a). It is noteworthy that $f_{\mathrm{infl}}$ also corresponds to the inflection point of $\Delta H_{\mathrm{Co}}$ (Fig.~\ref{fig:damping}(b)). The relative resonance intensities (see Fig.~\ref{fig:contour} and inset of Fig.~\ref{fig:IEC}) also change drastically around $f_{\mathrm{infl}}$. Both these effects are clear signs of a transition from a high-frequency region where the modes are associated with the individual Py and Co layers, to a low-frequency region where the precession is truly collective. $\Delta H_{\mathrm{Py}}$  decreases since it predominantly represents acoustic, in-phase, oscillations and the spin currents hence cancel, while they instead add up in the optical out-of-phase mode, resulting in an increasing $\Delta H_{\mathrm{Co}}$. When the Co/optical mode disappears, marked by arrows in Fig.~\ref{fig:damping}(a), the slope of $\Delta H_{\mathrm{Py}}$ again becomes constant and follows the single film behavior. The absence of spin-pumping enhanced damping implies that both layers precess in-phase and that the sum of the spin currents is virtually zero~\cite{Heinrich2003a}.

The Cu-40\AA~represents the samples without IEC and in those systems $\Delta H$ is linear in $f$, as expected for a pure spin pumping effect. The IEC not only influences the linewidth, but also the signal amplitude. The measured spectra (3-37GHz) of the Cu-40\AA~and Cu-8.8\AA~samples are presented in Fig.~\ref{fig:contour}(a) and~\ref{fig:contour}(b), respectively. The intensity of the signal is dependent on both the probed magnetic moment and the frequency. We have therefore normalized both resonances at each frequency to the highest amplitude. The Co mode of the Cu-40\AA~sample is stronger than the Py mode at all frequencies, as expected considering the higher magnetization and greater thickness of the Co layer. In contrast, the presence of IEC in the  Cu-8.8\AA~sample gives rise to a different picture, as its Co mode quickly decreases in amplitude below $\approx$~15 GHz. This reveals the transition to a region where the Py and Co show distinct acoustic and optical mode characteristics, in accordance with the interpretation of the frequency dependence of the linewidth and the shape of the $f_{\mathrm{r}}$ vs. $H_{\mathrm{r}}$ curves.

\begin{figure}[hbt]
\centering
\includegraphics[width=8cm]{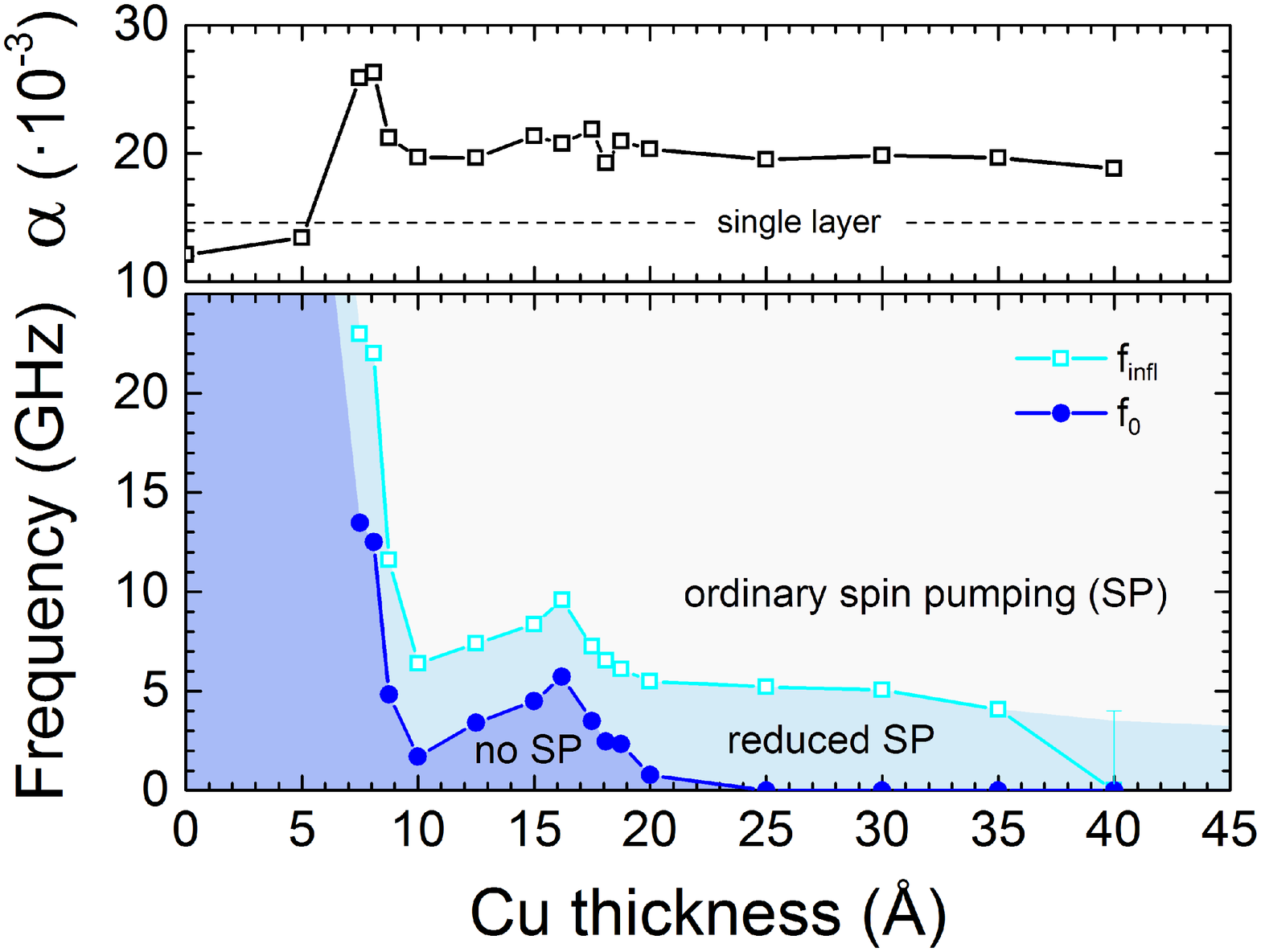}
\caption{Top) The damping parameter ($\alpha$) of the Py layers. Only data at frequencies well above $f_{\text{infl}}$ was used to extract $\alpha$. Bottom) The inflection point of the Co/optical-mode linewidth ($f_{\mathrm{infl}}$, cyan squares), which marks the transition from ordinary to reduced spin pumping, and the Co/optical-mode frequency at zero field ($f_{\mathrm{0}}$, blue circles), below which the spin pumping between the magnetic layers is absent.}
\label{fig:SP}
\end{figure}

The damping parameter $\alpha$ is determined by the relation $\Delta H ( f_{\mathrm{r}} ) = \Delta H_{0} + 4 \pi \alpha f_{\mathrm{r}}  / \gamma$, where $\Delta H_{0}$ is a zero-frequency offset~\cite{Heinrich2005book}. The top panel of Fig.~\ref{fig:SP} shows $\alpha_{\mathrm{Py}}$ extracted from $\Delta H$ in the linear region at high frequencies. The damping increases with IEC strength, as shown by the peaks around $t_{\mathrm{Cu}}=16$~and 8~\AA, since the precessing Py layer is dragged by the exchange field from the static Co layer. For even thinner $t_{\mathrm{Cu}}$ only the acoustic mode is present and $\alpha$ is low. Nonetheless, all samples showing two resonances has a higher damping compared to the single layer, and $\alpha$ is constant for the samples without coupling. Hence, the main source of the increased damping must be spin pumping.

The results are summarized in the bottom panel of Fig.~\ref{fig:SP}, which illustrates how the evolution of the resonances from Co and Py-like to truly acoustic and optical modes is associated with a reduction, and eventually a cancellation, of the spin pumping effects. Different frequency regimes are therefore characterized by dissimilar damping parameters and different levels of spin currents exchanged by the magnetic layers. This effect can be used to tailor the behavior of spintronic devices by the strength of the IEC. It is worth noting that the IEC is not only set within the growth process, but can also be tuned \emph{ex situ}, for example by loading the sample with hydrogen~\cite{Hjorvarsson1997}. It could therefore be possible to switch on and off the spin pumping in a device under operation.

In summary, we have investigated spin pumping in exchange coupled magnetic layers using broadband FMR spectroscopy. We observe a frequency dependence of the nature of the resonance modes. They display a single-layer like behavior at high frequencies and transform to characteristic collective modes as $f_{\mathrm{r}}$ decreases. This transition is accompanied by a reduction of the effective spin pumping and damping. The results demonstrate that it is possible to engineer a cut-off frequency, using the strength of the IEC, below which the spin pumping is minimized. 

We acknowledge financial support from ERC Starting Grant 307144 ``Mustang'', the Swedish Foundation for Strategic Research (SSF) Successful Research Leaders program, the Swedish Research Council (VR), the G\"{o}ran Gustafsson foundation, and the Knut and Alice Wallenberg Foundation.

\end{document}